\documentstyle{article}
\def\1ad{\mbox{\normalsize $^1$}}
\def\2ad{\mbox{\normalsize $^2$}}
\def\3ad{\mbox{\normalsize $^3$}}
\def\4ad{\mbox{\normalsize $^4$}}
\def\5ad{\mbox{\normalsize $^5$}}
\def\6ad{\mbox{\normalsize $^6$}}
\def\7ad{\mbox{\normalsize $^7$}}
\def\8ad{\mbox{\normalsize $^8$}}
\def\makefront{\vspace*{1cm}\begin{center}
\def\newtitleline{\\ \vskip 5pt}
{\Large\bf\titleline}\\
\vskip 1truecm
{\large\bf\authors}\\
\vskip 5truemm
\addresses
\end{center}
\vskip 1truecm
{\bf Abstract:}
\abstracttext
\vskip 1truecm}
\def\bfone{\relax{\rm 1\kern-.35em 1}}
\def\bfzero{\relax{\rm I\kern-.18em 0}}
\def\inbar{\vrule height1.5ex width.4pt depth0pt}
\def\IC{\relax\,\hbox{$\inbar\kern-.3em{\rm C}$}}
\def\ID{\relax{\rm I\kern-.18em D}}
\def\IF{\relax{\rm I\kern-.18em F}}
\def\IK{\relax{\rm I\kern-.18em K}}
\def\IH{\relax{\rm I\kern-.18em H}}
\def\II{\relax{\rm I\kern-.17em I}}
\def\IN{\relax{\rm I\kern-.18em N}}
\def\IP{\relax{\rm I\kern-.18em P}}
\def\IQ{\relax\,\hbox{$\inbar\kern-.3em{\rm Q}$}}
\def\IR{\relax{\rm I\kern-.18em R}}
\def\IG{\relax\,\hbox{$\inbar\kern-.3em{\rm G}$}}
\font\cmss=cmss10 \font\cmsss=cmss10 at 7pt
\def\ZZ{\relax\ifmmode\mathchoice
{\hbox{\cmss Z\kern-.4em Z}}{\hbox{\cmss Z\kern-.4em Z}}
{\lower.9pt\hbox{\cmsss Z\kern-.4em Z}}
{\lower1.2pt\hbox{\cmsss Z\kern-.4em Z}}\else{\cmss Z\kern-.4em
Z}\fi}

 \def\cM{{\cal M}}
\def\cN{{\cal N}} 
\def\cP{{\cal P}} 
 
\def\bar{\overline}

\def\Coe#1.#2.{{#1\over #2}}

\def\coe#1.#2.{\relax{\textstyle {#1 \over #2}}\displaystyle}

\def\to{\rightarrow}
\def\notin{\hbox{{$\in$}\kern-.51em\hbox{/}}}

\def\IE{\relax{{\rm I\kern-.18em E}}}

\def\IGam{\relax{{\rm I}\kern-.18em \Gamma}}

\def\IA{\relax{\hbox{{\rm A}\kern-.82em {\rm A}}}}

\newcommand{\be}{\begin{equation}}
\newcommand{\ee}{\end{equation}}
\newcommand{\ba}{\begin{eqnarray}}
\newcommand{\ea}{\end{eqnarray}}
\newtheorem{definizione}{Definition}[section]
\newcommand{\bd}{\begin{definizione}}
\newcommand{\ed}{\end{definizione}}
\newtheorem{teorema}{Theorem}[section]
\newcommand{\bth}{\begin{teorema}}
\newcommand{\eth}{\end{teorema}}
\newtheorem{lemma}{Lemma}[section]
\newcommand{\blem}{\begin{lemma}}
\newcommand{\elem}{\end{lemma}}
\newcommand{\brr}{\begin{array}}
\newcommand{\err}{\end{array}}
\newcommand{\nn}{\nonumber}
\newtheorem{corollario}{Corollary}[section]
\newcommand{\bcorol}{\begin{corollario}}
\newcommand{\ecorol}{\end{corollario}}

\begin {document}
\def\titleline{
Central charges, black-hole entropy and geometrical structure
\newtitleline
of $N$--extended supergravities in $D=4$
}
\def\authors{Laura Andrianopoli
}
\def\addresses{Dipartimento di Fisica, Universit\`a di Genova,
Via Dodecaneso, 33, I-16146 Genova, Italy\\
and Istituto Nazionale di Fisica Nucleare (I.N.F.N.) - Sezione di
Torino, Italy
}
\def\abstracttext{
The derivation of absolute  (moduli-independent)  U-invariants  for
all  $N>2$ extended
supergravities   at  $D=4$    in  terms  of  (moduli-dependent)
central and matter charges is reported.
\\
These  invariants  give a general definition of the ``topological''
Bekenstein--Hawking  entropy  formula for extremal black-holes and reduce to
the  square  of the black-hole ADM mass for ``fixed scalars'' which extremize
the black-hole ``potential'' energy.

}
\makefront
\section{Introduction}
Recently, considerable progress has been made in the study of general
properties
of black holes arising in supersymmetric theories of gravity such as  extended
supergravities, string theory and M-theory \cite{string}.
Of particular interest are extremal black holes in four dimensions which
correspond to BPS saturated states \cite{black} and whose ADM mass depends,
 beyond the
quantized values of electric and magnetic charges, on the asymptotic value
 of scalars at infinity.
The latter describe the moduli space of the theory.

Another physical relevant quantity, which depends only on quantized electric
 and magnetic charges,
 is the black hole entropy,
which can be defined macroscopically, through the Bekenstein-Hawking
 area-entropy relation
or microscopically, through D-branes techniques \cite{dbr} by counting
of microstates \cite{micros}.
It has been further realized that the scalar fields, independently of
their values
at infinity, flow towards the black hole horizon to a fixed value of pure
 topological
nature given by a certain ratio of electric and magnetic charges \cite{fks}.

These ``fixed scalars'' correspond to the extrema of the ADM mass
in moduli space while the black-hole entropy  is the actual value of the
 squared
ADM mass at this point \cite{feka1}.

In theories with $N>2$, extremal black-holes preserving one supersymmetry
have
the further property that all central charge eigenvalues other than the one
equal to the BPS mass flow to zero for ``fixed scalars''.
The black-hole entropy is still given by the square of the ADM mass for
 ``fixed scalars''\cite{feka2}.

Recently \cite{fegika}, the nature of these extrema has been further studied
and shown that  they
generically correspond to non degenerate minima for $N=2$ theories whose
 relevant
moduli space is the special geometry of $N=2$ vector multiplets.

The entropy formula turns out to be in all cases a U-duality invariant
expression
(homogeneous of degree two) built out of electric and magnetic charges and as
 such
can be in fact also computed through certain (moduli-independent) topological
quantities which only depend on the nature of the U-duality groups and the
appropriate representations
of electric and magnetic charges.
For example, in the $N=8$ theory  the entropy was shown to correspond  to the
unique quartic $E_7$ invariant built with its 56 dimensional representation
\cite{kall}.

In this talk we
intend to report on further progress made on this subject in collaboration
with Riccardo D'Auria and Sergio Ferrara \cite{marzo}
 by  deriving, for all $N>2$
theories,
topological (moduli-independent) U-invariants constructed in terms of
(moduli-dependent)
 central charges and matter charges, and show that, as expected, they coincide
with the squared ADM mass at ``fixed scalars''.

\section{Central charges, U-invariants and entropy}
Extremal   black-holes   preserving   one   supersymmetry  correspond  to
$N$-extended multiplets with
\begin{equation}
M_{ADM} = \vert Z_1 \vert >  \vert Z_2 \vert  \cdots > \vert Z_{[N/2]} \vert
\end{equation}
where $Z_\alpha$, $\alpha =1,\cdots, [N/2]$, are the proper values of
 the central charge antisymmetric matrix written in normal form
 \cite{fesazu}.
The central charges $Z_{AB}= -Z_{BA}$, $A,B=1,\cdots,N$, and matter charges
 $Z_I$, $I= 1,\cdots , n$ are
those (moduli-dependent) symplectic invariant combinations of field strenghts
and their duals
(integrated over a large two-sphere)
 which appear
in the gravitino and gaugino supersymmetry variations respectively
\cite{cedafe}, \cite{noi1}, \cite{noi}.
Note that the total number of vector fields is $n_v=N(N-1)/2+n$ (with the
 exception of $N=6$
in which case there is an extra singlet graviphoton)\cite{cj}.
         \\
  It was shown in ref. \cite{feka2} that at the attractor point, where
  $M_{ADM}$ is extremized, supersymmetry requires that $Z_\alpha$, $\alpha >1$,
  vanish toghether with the matter charges $Z_I$, $I= 1, \cdots , n$
($n$ is the number of matter multiplets, which can exist only for $N=3,4$)
\par
Moreover, S. Ferrara, G. Gibbons, R. Kallosh and B. Kol showed that
the properties of extreme non rotating black holes are
 completely encoded in the metric of the manifold $G_{ij}$  spanned by
 scalar fields
and on an effective potential, function of scalars and electric and
magnetic charges, $V(\Phi, e,g)$, known as
{\it geodesic potential} \cite{bmgk}\cite{fegika}, defined by:
\begin{equation}
V=-{1\over 2}P^t\cM(\cN)P.
\label{sumrule1}
\end{equation}
Here $P$ is the symplectic vector $P=(p^\Lambda, q_\Lambda) $ of quantized
electric and
magnetic charges and $\cM(\cN)$ is a symplectic $2n_v \times 2n_v$ matrix
 whose
$n_v\times n_v$ blocks are given in terms of the $n_v\times n_v$ vector
 kinetic matrix $\cN_{\Lambda\Sigma}$
($-Im \cN, Re \cN$ are the normalizations of the kinetic $F^2$ and the
topological $F^*F$ terms respectively appearing
in the black-hole supergravity lagrangian) and
\begin{equation}
\cM(\cN) = \pmatrix{A& B \cr C & D \cr}
\label{mn}
\end{equation}
with:
\begin{eqnarray}
A&=& Im \cN + Re \cN Im \cN^{-1} Re\cN \nonumber\\
B&=& - Re \cN Im \cN^{-1} \nonumber\\
C&=& -  Im \cN^{-1} Re\cN \nonumber\\
D&=&Im \cN^{-1}
\end{eqnarray}
In particular, they showed \cite{bmgk}\cite{fegika} that the area
of the event horizon is proportional to the value of $V$ at
the horizon:
\be
{A\over 4\pi}=V(\Phi_h,e,g)
\ee
where $\Phi_h$ is the value taken by scalar fields at the horizon.
It was also shown that in a general (even not supersymmetric) setting,
Einstein equations can be reduced to a system of scalar equations
that give, near the horizon, the solution:
\be
\Phi^i = \left({2\pi \over A}\right){\partial V \over \partial \Phi^i}
        \mbox{log}\tau + \Phi^i_h.
        \label{phih}
\ee
in terms of the evolution parameter $\tau = \frac{1}{r-r_h}$.
From eq. (\ref{phih}) we see that the request that the horizon is a fixed
point (${d\Phi^i \over d \tau}=0$) implies that the geodesic potential
is extremized in moduli space:
\be
\Phi_h \, : \quad {d\Phi^i \over d \tau}=0 \quad \leftrightarrow \quad
{\partial V \over \partial \Phi^i}|_{\Phi_h} =0
\ee
To summarize,  the area $A_h$  of the event horizon,
and then, from the area--entropy formula $S_{B-H}={A_h\over 4}$,
the Bekenstein--Hawking entropy $S_{B-H}$ of the black hole, are given
by the geodesic potential evaluated at the horizon, and we have
a tool for finding this value:
the geodesic potential gets an extremum at the horizon.
\par
However, the geodesic potential $V(\Phi, e,g)$ defined in eq.s (\ref{sumrule1})
and (\ref{mn}) has a particular meaning in supergravity theories,
that allows to find its extremum in a very easy way.
Indeed, an expression exactly coinciding with (\ref{sumrule1}) has been found
in \cite{noi} as the result of
a sum rule among central and matter charges in supergravity theories.
So, in every supergravity theory, the geodesic potential has the
general form:
\be
V\equiv -{1\over 2}P^t\cM(\cN)P ={1\over 2} Z_{AB}\bar Z^{AB}+Z_I \bar Z^I
\ee
Central and matter charges in extended supergravities
were found to satisfy some differential relations \cite{noi},
that are a direct consequence of the
geometrical properties of the manifolds spanned by scalar fields, in
particular the fact of being coset manifolds of the form $U/H$, embedded
in the symplectic group $Sp(2n_v)$ ($n_v$ being the total number of vectors)
\cite{noi}.

Then, to find the extremum of $V$ we can apply the differential relations
among central and matter charges found in \cite{noi},
that in general read:
\begin{eqnarray}
  \nabla Z_{AB} &=&  \bar Z_{I}  P^I_{AB} +  {1 \over 2} \bar Z^{CD} P_{ABCD} \nonumber \\
\nabla Z_{I} &=& {1 \over 2}  \bar Z^{AB}  P_{AB I} + \bar Z^{J}  P_{JI}
 \label{diffcharge}
\end{eqnarray}
where the matrices $P_{ABCD}$, $P_{AB I}$, $P_{IJ}$ are the subblocks of the
 vielbein of $U/H$ \cite{noi}:
\begin{equation}
  \label{vielbein}
  \cP \equiv L^{-1} \nabla L = \pmatrix{P_{ABCD}& P_{AB I} \cr
P_{I AB} & P_{IJ} \cr }
\end{equation}
written in terms of the indices of $H=H_{Aut} \times H_{matter}$.

Applying eq.s (\ref{diffcharge}) to the geodesic potential, we find that
the extremum is given by:
\ba
dV &=& {1\over 2}\nabla Z_{AB}\bar Z^{AB}+\nabla Z_I \bar Z^I  +c.c. =\nn\\
   &=& {1\over 2}\left({1 \over 2} \bar Z^{CD} P_{ABCD}+
        \bar Z_{I}  P^I_{AB} \right)\bar Z^{AB} \nn\\
   &+& \left({1 \over 2}  \bar Z^{AB}  P_{AB I} + \bar Z^{J}  P_{JI}\right)
        \bar Z^I =0
\ea
that is $dV=0$ for:
\be
Z_I=0 \, ; \, \bar Z^{AB} \bar Z^{CD} P_{ABCD} =0
\label{extgeo}
\ee

 However, in the following we will show one more technique for finding
 the entropy, exploiting the fact that it is a `topological quantity' not
 depending on scalars. This last procedure is particularly interesting
 because it refers only to group theoretical properties of the coset manifolds
 spanned by scalars, and do not need the knowledge of any details of the
 black-hole horizon.
 This allows to give the
entropy formula as a moduli--independent quantity in the entire
moduli space and not just at the critical points.
Namely, we are looking for quantities $S\left(Z_{AB}(\phi), \bar Z^{AB}
 (\phi),Z_{I}(\phi), \bar Z^{I} (\phi)\right)$
such that ${\partial \over \partial \phi ^i} S =0$, $\phi ^i$ being the moduli
 coordinates
 \footnote{The Bekenstein-Hawking entropy $S_{BH} ={A\over 4}$ is actually
$\pi S$
in our notation.}.
\\
These formulae generalize the quartic $E_{7(-7)}$ invariant of $N=8$
supergravity \cite{kall} to all other cases.
\par
Let us first consider theories  where  matter can be
present ($N=3,4$)\cite{maina}, \cite{bks}.
We focus in particular on the
$N=4$ case.
\par
The U--duality group \footnote{Here we denote by U-duality group
the isometry group $G$
acting on the scalars, although only a restriction of it to integers is the
 proper U-duality group \cite{ht}.}
 is, in this case, $SU(1,1)
\times SO(6,n)$.
The central and matter charges $Z_{AB}, Z_I$ transform under the
isotropy group:
\begin{equation}
 H= SU(4) \times O(n) \times U(1)
\end{equation}
Under the action of the elements of $U/H$ the charges get mixed with
their complex conjugate. The transformation
of the charges under infinitesimal transformations $\xi$ on the coset
 $K= {SU(1,1) \over U(1)} \times
{O(6,n) \over O(6) \times O(n)}$
can be read from the differential relations (\ref{diffcharge})
 satisfied by the charges
 \cite{noi}:
 \begin{eqnarray}
\delta Z_{AB}  &=& {1\over 2} \xi  \epsilon_{ABCD} \bar Z^{CD}  +
 \xi _{AB I} \bar Z^I \\
\delta Z_{I}  &=& \xi \eta_{IJ} \bar Z^J + {1\over 2}\xi _{AB I} \bar Z^{AB}
\label{deltaz4}
\end{eqnarray}
where we used the fact that for the  $N=4$  theory the vielbein components
have the form:
\begin{equation}
P_{ABCD} = \epsilon_{ABCD}P ,\quad P_{IJ} = \eta_{IJ}P,\quad
P_{AB I}={1\over 2} \eta_{IJ} \epsilon_{ABCD}\bar P^{CD J},
\label{viel4}
\end{equation}
and that, once the covariant derivatives (\ref{diffcharge}) are known,
the infinitesimal variations are simply
obtained by the substitution $\nabla \to \delta$, $P \to \xi$
(with $\bar \xi^{AB I} =  {1\over 2} \eta^{IJ}   \epsilon^{ABCD}
\xi_{CD J}$).
\par
There are three $O(6,n)$ invariants given by $I_1$, $I_2$, $\bar I_2$ where:
\begin{eqnarray}
  I_1 &=& {1 \over 2}  Z_{AB} \bar Z_{AB} - Z_I \bar Z^I
\label{invar41} \\
I_2 &=& {1\over 4} \epsilon^{ABCD}  Z_{AB}   Z_{CD} - \bar Z_I \bar Z^I
\label{invar42}
\end{eqnarray}
and the unique   $SU(1,1)  \times
O(6,n) $  invariant $S$, $\nabla S =0$, is given by:
\begin{equation}
S= \sqrt{(I_1)^2 - \vert I_2 \vert ^2 }
\label{invar4}
\end{equation}

At the attractor point $Z_I =0 $ and $\epsilon^{ABCD} Z_{AB} Z_{CD}
=0$ so that $S$ reduces to the square of the BPS mass.
\par
For $N=5,6,8$ the U-duality invariant expression $S$ is the square
root of a unique invariant under the corresponding U-duality groups
$SU(5,1)$, $O^*(12)$ and $E_{7(-7)}$.
The strategy is to find a quartic expression $S^2$    in terms of
$Z_{AB}$ such that $\nabla S=0$, i.e. $S$ is moduli-independent.
\par
As before, this quantity is a particular combination of the $H$
quartic invariants.
\par
Let us analyse in particular the $N=8$ case.
For $N=8$ the $SU(8)$ invariants are \footnote{The Pfaffian of an
 $(n\times n)$ ($n$ even) antisymmetric
matrix is defined as $Pf Z={1\over 2^n n!} \epsilon^{A_1 \cdots A_n}
 Z_{A_1A_2}\cdots Z_{A_{N-1}A_N}$, with the property:
$ \vert Pf Z \vert = \vert det Z \vert ^{1/2}$. }:
\begin{eqnarray}
I_1 &=& \left(Z_{AB}\bar Z^{BA}\right)^2 \equiv  (Tr A) ^2 \\
I_2 &=& Z_{AB} \bar Z^{BC} Z_{CD} \bar Z^{DA} \equiv Tr (A^2) \\
I_3 &=& Pf \, Z
 ={1\over 2^4 4!} \epsilon^{ABCDEFGH} Z_{AB} Z_{CD} Z_{EF} Z_{GH}
\end{eqnarray}
where $A_A^{\ B} \equiv Z_{AC}\bar Z^{CB}$.
\par
The ${E_{7(-7)} \over SU(8)}$ transformations are:
\begin{equation}
\delta Z_{AB} ={1\over 2} \xi_{ABCD} \bar Z^{CD}
\end{equation}
where $\xi_{ABCD}$ satisfies the reality constraint:
\begin{equation}
\xi_{ABCD} = {1 \over 24} \epsilon_{ABCDEFGH} \bar \xi^{EFGH}
\end{equation}
One finds the following $E_{7(-7)}$ invariant \cite{kall}:
\begin{equation}
S= {1\over 2} \sqrt{4 Tr (A^2) - ( Tr A)^2 + 32 Re (Pf \,
Z) }
\end{equation}
Note that at the attractor point $Pf\,Z =0$, $(Tr A)^2=2Tr (A^2)$
and $S$ becomes proportional to the square of the BPS mass.

\section*{Acknowledgements}
The work reported in this talk was performed in collaboration with
Riccardo D'Auria and Sergio Ferrara.




\begin{thebibliography}{99}
\bibitem{string}
For a review, see for instance:
M. J. Duff, R. R. Khuri and J. X. Lu, {\it String solitons}, Phys. Rep.
 {\bf 259} (1995) 213;
M. J. Duff, Kaluza-Klein theory in perspective, in {\it Proceedings of
 the Nobel Symposium Oskar
Klein Centenary, Stockholm, September 1994} (World Scientific, 1995),
 E. Lindstrom editor, hep-th/9410046;
G. Horowitz, UCSBTH-96-07, gr-qc/9604051;
J. M. Maldacena, Ph.D. thesis, hep-th/9607235;
M. Cvetic, UPR-714-T, hep-th/9701152
\bibitem{black}
G. Gibbons, in {\it Unified theories of Elementary Particles. Critical
Assessment
and Prospects}, Proceedings of the Heisemberg Symposium, M\"unchen, West
 Germany, 1981,
 ed. by P. Breitenlohner and H. P. D\"urr, Lecture Notes in Physics Vol.
 160 (Springer-Verlag, Berlin, 1982);
G. W. Gibbons and C. M. Hull, Phys. lett. {\bf 109B} (1982) 190;
G. W. Gibbons, in {\it Supersymmetry, Supergravity and Related Topics},
 Proceedings
of the XVth GIFT International Physics, Girona, Spain, 1984, ed. by F. del
 Aguila, J.
de Azc\'arraga and L. Ib\'a\~nez, (World Scientific, 1995), pag. 147;
R. Kallosh, A. Linde, T. Ortin, A. Peet and A. Van Proeyen, Phys. Rev.
 {\bf D46} (1992) 5278;
R. Kallosh, T. Ortin and A. Peet, Phys. Rev. {\bf D47} (1993) 5400;
R. Kallosh, Phys. Lett. {\bf B282} (1992) 80;
R. Kallosh and A. Peet, Phys. Rev. {\bf D46} (1992) 5223;
A. Sen, Nucl. Phys. {\bf B440} (1995) 421; Phys. Lett. {\bf B303} (1993) 221;
 Mod. Phys. Lett. {\bf A10}
(1995) 2081;
J. Schwarz and A. Sen, Phys. Lett. {\bf B312} (1993) 105;
M. Cvetic and D. Youm, Phys. Rev. {\bf D53} (1996) 584;
M. Cvetic and A. A. Tseytlin, Phys. Rev. {\bf D53} (1996) 5619;
M. Cvetic and C. M. Hull, Nucl. Phys. {\bf B480} (1996) 296
\bibitem{dbr}
A. Strominger and C. Vafa, Phys. Lett. {\bf B379} (1996) 99, hep-th/9601029;
C. G. Callan and J. M. Maldacena, Nucl. Phys. {\bf B472}
 (1996) 591, hep-th/9602043;
G. Horowitz and A. Strominger, Phys. Rev. Lett. {\bf B383} (1996) 2368,
 hep-th/9602051;
R. Dijkgraaf, E. Verlinde, H. Verlinde, Nucl.Phys. {\bf B486} (1997) 77,
 hep-th/9603126;
P. M. Kaplan, D. A. Lowe, J. M. Maldacena and A. Strominger, hep-th/9609204;
 J. M. Maldacena, hep-th/9611163
\bibitem{micros}
L. Susskind, hep-th/9309145; L. Susskind and J. Uglum, Phys. Rev. {\bf D50}
 (1994) 2700;
F. Larsen and F. Wilczek, Phys. Lett. B375 (1996) 37, hep-th/9511064
\bibitem{fks}
S. Ferrara, R. Kallosh and A. Strominger, Phys. Rev. {\bf D52} (1995) 5412,
 hep-th/9508072;
A. Strominger, Phys. Lett. {\bf B383} (1996) 39, hep-th/9602111
\bibitem{feka1}
S. Ferrara and R. Kallosh, Phys. Rev. {\bf D54} (1996) 1514, hep-th/9602136
\bibitem{feka2}
S. Ferrara and R. Kallosh, Phys. Rev. {\bf D54} (1996) 1525, hep-th/9603090
\bibitem{fegika}
S. Ferrara, G. W. Gibbons and R. Kallosh, hep-th/9702103
\bibitem{kall}
R. Kallosh and B. Kol, Phys. Rev. {\bf D53} (1996) 5344
\bibitem{marzo}
L. Andrianopoli, R. D'Auria and S. Ferrara, Phys. Lett. {\bf B403} (1997) 12,
hep-th/9703156
\bibitem{fesazu}
S. Ferrara, C. Savoy and B. Zumino, Phys. Lett. {\bf 100B} (1981) 393
\bibitem{cedafe}
A. Ceresole,  R. D'Auria and S. Ferrara, in ``{\it S-Duality and Mirror
 symmetry}'',
Nucl. Phys. (Proc. Suppl.) {\bf B46} (1996) 67, ed. E. Gava, K. S. Narain
and C. Vafa,
hep-th/9509160
\bibitem{noi1}
L. Andrianopoli, R. D'Auria and S. Ferrara, hep-th/9608015,
Int. Jour. Mod. Phys. {\bf A 12}(1997) 3759
\bibitem{noi}
L. Andrianopoli, R. D'Auria and S. Ferrara, hep-th/9612105
to appear in
 International Journal of Modern Physics A
 \bibitem{cj}
E. Cremmer in ``{\it Supergravity '81}'', ed. by S. Ferrara and J. G. Taylor,
 Pag. 313;
B. Julia in ``{\it Superspace \& Supergravity}'', ed. by S. Hawking and M.
 Rocek, Cambridge (1981) pag. 331
\bibitem{bmgk}
P. Breitenlohner, D. Maison and G. W. Gibbons, Commun. Math. Phys. {\bf 120}
 (1988) 295;
G. W. Gibbons, R. Kallosh and B. Kol, Phys. Rev. Lett. {\bf 77} (1996) 4992,
 hep-th/9607108

\bibitem{ht}
C. M. Hull and P. K. Townsend, Nucl. Phys.
hep-th/9410167, Nucl. Phys. {\bf B438} (1995) 109

\bibitem{maina}
L. Castellani, A. Ceresole, R. D'Auria, S. Ferrara, P. Fr\'e and E. Maina,
 Nucl. Phys. {\bf B286} (1986)
317
\bibitem{bks}
E. Bergshoeff, I. G. Koh and E. Sezgin, Phys. Lett. {\bf 155B} (1985) 71;
M. de Roo and F. Wagemans, nucl. Phys. {\bf B262} (1985) 644

\end{thebibliography}
\end{document}